\documentclass[prb,aps,twocolumn]{revtex4}
\input epsf
\input pstricks
\begin{document}
\title{On the premelting features in sodium clusters}
\author{F. Calvo and F. Spiegelman}
\affiliation{Laboratoire de Physique Quantique, IRSAMC, Universit\'e Paul
Sabatier, 118 Route de Narbonne, F31062 Toulouse Cedex, France}
\begin{abstract}
Melting in Na$_n$ clusters described with an empirical embedded-atom
potential has been reexamined in the size range $55\leq n\leq 147$
with a special attention at sizes close to 130. Contrary to previous
findings, premelting effects are also present at such medium sizes,
and they turn out to be even stronger than the melting process itself
for Na$_{133}$ or Na$_{135}$. These results indicate that the
empirical potential is {\em qualitatively}\/ unadequate to model
sodium clusters.
\end{abstract}
\maketitle

\section*{Introduction}

The early days of cluster thermodynamics were first mostly concerned with
theoretical of numerical studies of simple rare-gas aggregates. Beyond
the now generally accepted idea that melting in finite atomic clusters
appears as a first-order transition rounded by size effects,\cite{labwhet}
it was also observed in simulations that these system can exhibit
dynamical coexistence,\cite{berry} a process in which the cluster
fluctuates in time between its solidlike and liquidlike states.

While no experiment has yet permitted to test these predictions on the
very same clusters, recent observations achieved in the group lead by
Haberland\cite{hab1,hab2,hab3} have provided valuable qualitative and
quantitative information about the way sodium clusters melt, by
extracting the full caloric curves from photodissociation
measurements. In particular these authors reported that the melting
point and the latent heat of fusion both vary strongly and
nonmontonically with size.\cite{hab2} These results significantly
attracted the attention of theoreticians, who since then tried to interpret
or reproduce these unexpected complex variations using a variety of
models.\cite{cs1,cs2,cs3,aguado,manninen1,manninen2,garzon}
Further evidence was also seeked in indicators, which are alternative
to the caloric curves, such as the electric
polarizability,\cite{kummel} the photoabsorption
spectrum,\cite{moseler} or the ionization potential.\cite{manninen2}

Up to now, none of the above cited theoretical works has been able
to reach a fully satisfactory quantitative agreement with experiments.
The situation is largely due to the expected strong interplay between
geometric and electronic effects, which could be responsible for the
variations of the thermodynamic quantities. However, the previous
simulations have brought some clues about the relevance of the various
models and potentials used to describe simple metal clusters. For
instance, it was seen that the distance dependent tight-binding (TB)
model developed by Poteau and Spiegelman\cite{poteau} overestimated
melting points by more than 20 percents. This was interpreted as the
consequence of the parameterization of this model, carried out on
small clusters only but not on bulk properties.\cite{cs2}
On the other hand, both the empirical embedded-atom model (EAM)
potential\cite{cs1,cs2,manninen2,garzon} and orbital-free density
functional\cite{aguado} calculations lead to a notable
underestimation. Unfortunately, more realistic simulations still lack
the extent of phase space sampling required for a precise computation
of the melting point above 50 atoms. 

\section*{Numerical experiments}

In our previous works, we concluded that melting in sodium clusters
actually occurs differently in the smallest and in medium to large
sizes clusters.\cite{cs1,cs2} While the caloric curve of the smallest
clusters (having less than about 80 atoms) exhibit several features
due to multiple isomerizations prior melting, the heat capacity of
the larger sizes mainly has one main peak indicating melting in a
non ambiguous way. This could partly explain why experimental
measurements do not find such premelting features above 55 atoms,
and also why they have difficulties in getting a clear, single-peak
picture below this size. More recently we noticed that premelting
effects could also be artefactually due to poorly converged
simulations.\cite{cs3}

Based on our previous results, the disagreement between
experimental measurements and the results of both the TB and EAM
models seemed mostly quantitative. One could hope in getting a much
better agreement by suitably modifying the parameters, after including
both molecular and bulk properties, possibly through allowing these
parameters to become size-dependent. However, the range of sizes that
was investigated by us and by others was quite
limited,\cite{cs1,cs2,manninen2,garzon} with only very few sizes above
80. Moreover, the mediocre agreement for the latent heats of fusion
lead us revisit the problem with newly available simulation methods
and less ambiguous tools of analysis.

We have performed exchange Monte Carlo
canonical simulations\cite{emc} of the
clusters Na$_n$ with $55\leq n\leq 145$, $n$ being a multiple of 5 in
this range, plus the following sizes $n=59$, 93, 127, 129--131, 133,
139, 142, and 147. The clusters are again described using the same
empirical many-body EAM potential whose parameters are given in Ref.
\onlinecite{li}, and each simulation consisted of $10^7$ cycles
following $2\times 10^6$ equilibration cycles for each of the 31
trajectories characterized by their temperature $T_i=i\times 10$~K
for $1\leq i\leq 30$ plus $i=1.5$. The starting structures for all
clusters was always chosen to be the result of basin-hopping global
optimization carried out with 10 sets of $10^4$ quenches. All were
found to be based on the icosahedral motif. The absence of any
different structural motif such as octahedral or decahedral for the
sizes studied here is very favorable for the simulations to reach
equilibrium and not fall into broken ergodicity problems, as one
major cause for such difficulties precisely lies in the energy
landscape having several funnels.\cite{wales38,neirotti}
Each cluster was simulated 5 times independently, with different
random seeds, and we used a hard-wall spherical container with radius
$R_{\rm max} = 7n^{1/3}$. The caloric curves were constructed from the
distributions of potential energies using a multihistogram technique.
In Ref. \onlinecite{cs2}, we calculated the latent heat of melting,
$L$, as the integral of the heat capacity minus the Dulong-Petit
contribution. This lead to appreciable overestimates, mostly due to
the neglect of anharmonicities. Here we proceed similarly to the
experimental approach of Schmidt and coworkers,\cite{hab2} namely by
fitting the low- and high-temperature parts of the internal energy as
straight lines and defining $L$ as the gap between these lines at the
melting point $T_{\rm melt}$. The melting point itself is defined as
the temperature at which the {\em last}\/ heat capacity peak has its
maximum: in cases where there are several peaks each centered around
$T_{\rm melt}^{(k)}$ the true melting temperature is taken as
$\max_k\{ T_{\rm melt}^{(k)} \}$. The low- and high-temperature parts
are defined as $T\leq 50$~K and $T\geq 250$~K,
respectively. Therefore, if premelting events are present between
these limits, they will contribute to the latent heat.

\section*{Results and discussion}

\begin{figure}[htb]
\setlength{\epsfxsize}{9cm} \centerline{\epsffile{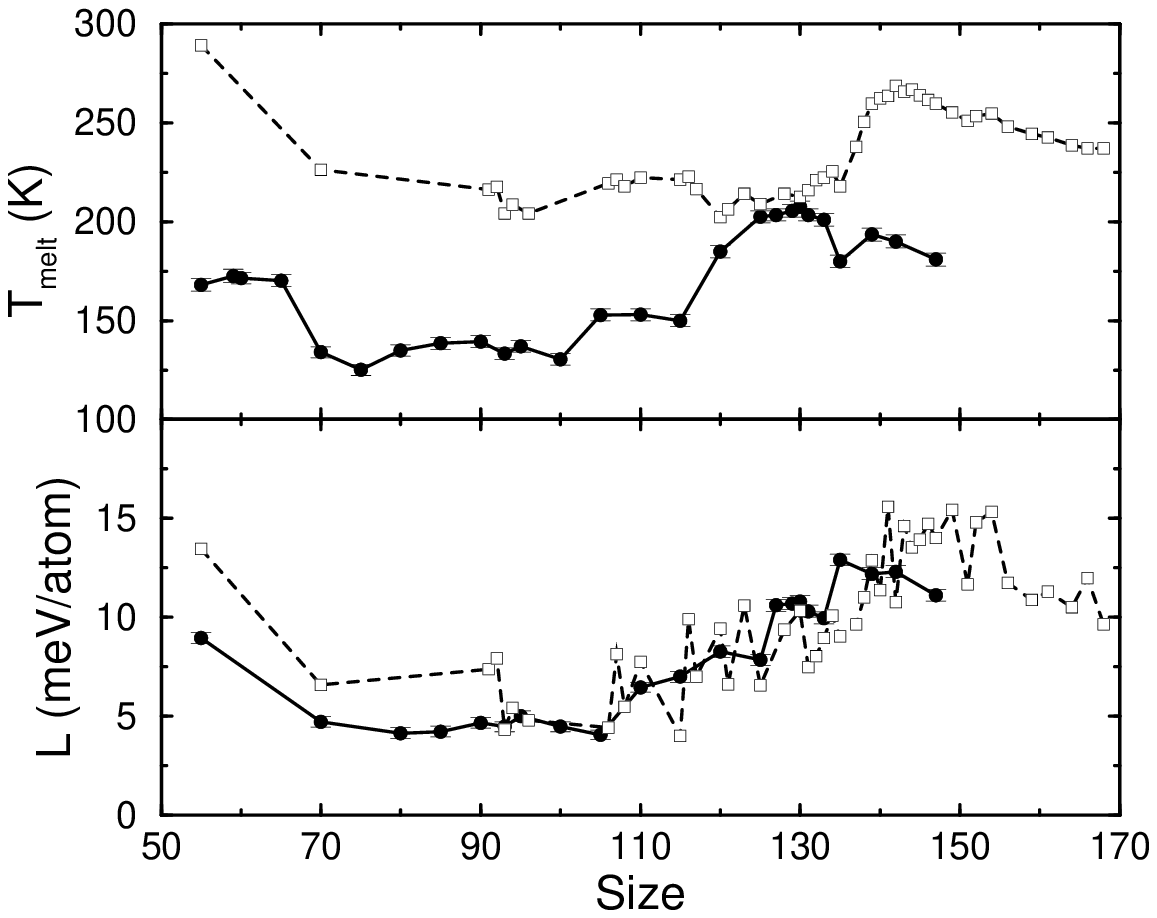}}
\caption{Melting point $T_{\rm melt}$ and latent heat of melting $L$
of sodium cluster clusters versus their size. The open squares are the
experimental results of Schmidt {\em et al.}\protect\cite{hab3}, the
full circles are for Monte Carlo simulations using the empirical
potential.}
\label{fig:compexp}
\end{figure}

The variations of $T_{\rm melt}$ and $L$ with size are sketched in
Fig.~\ref{fig:compexp} along with the latest experimental data of
Schmidt and coworkers.\cite{hab3} Except for very low sizes, we first
notice that the melting points computed here are significantly lower
than the ones we previously reported.\cite{cs2} This is an unfortunate
consequence of using the q-jumping method with un inappropriate Tsallis
parameter.\cite{difres} However they are comparable to the ones obtained
by Garz\'on and coworkers in the microcanonical ensemble.\cite{garzon}
Since we expect the differences between this ensemble and the
canonical ensemble to become smaller and smaller for increasing sizes,
this agreement appear as a mutual confirmation of the convergence of
both calculations, even though they rely on very different numerical
experiments.

Interestingly, the experimental and theoretical variations of
$T_{\rm melt}$ with size appear to be related to each other,
especially if we shift the simulation data around size 120 by about
15 atoms and 70~K. This had not been noticed before, and is probably
fortuitous. But it may also hide that some mechanisms causing the strong
variations in the experimental results close to 140 atoms are indeed
the same here, only occuring sooner.

In Fig.~\ref{fig:compexp} one should also notice that the melting
point at $n=55$ is not the highest, as clusters having 59 or 60 atoms
are more resistant to an increase in temperature. However, we were not
able to extract any latent heat for these sizes (nor for $n=65$ and
70) due to very broad heat capacity peaks. Therefore one should maybe
not give too much importance to the melting points extracted from
these curves.

While the melting temperatures are usually well below the experimental
data (except in the vicinity of 130 atoms), the computed latent heats
show a reasonable overall agreement, even though the complex
variations are not as sharply seen as in the measurements of Schmidt
{\em et al.}\cite{cs3} A definitive comparison would require one to
extend the range of sizes. As far as latent heats are concerned,
the difference between the present results and our previously
published data\cite{cs2} comes nearly entirely from the different way
of estimating $L$, which is now much closer to the experimental way.

\begin{figure}[htb]
\setlength{\epsfxsize}{9cm} \centerline{\epsffile{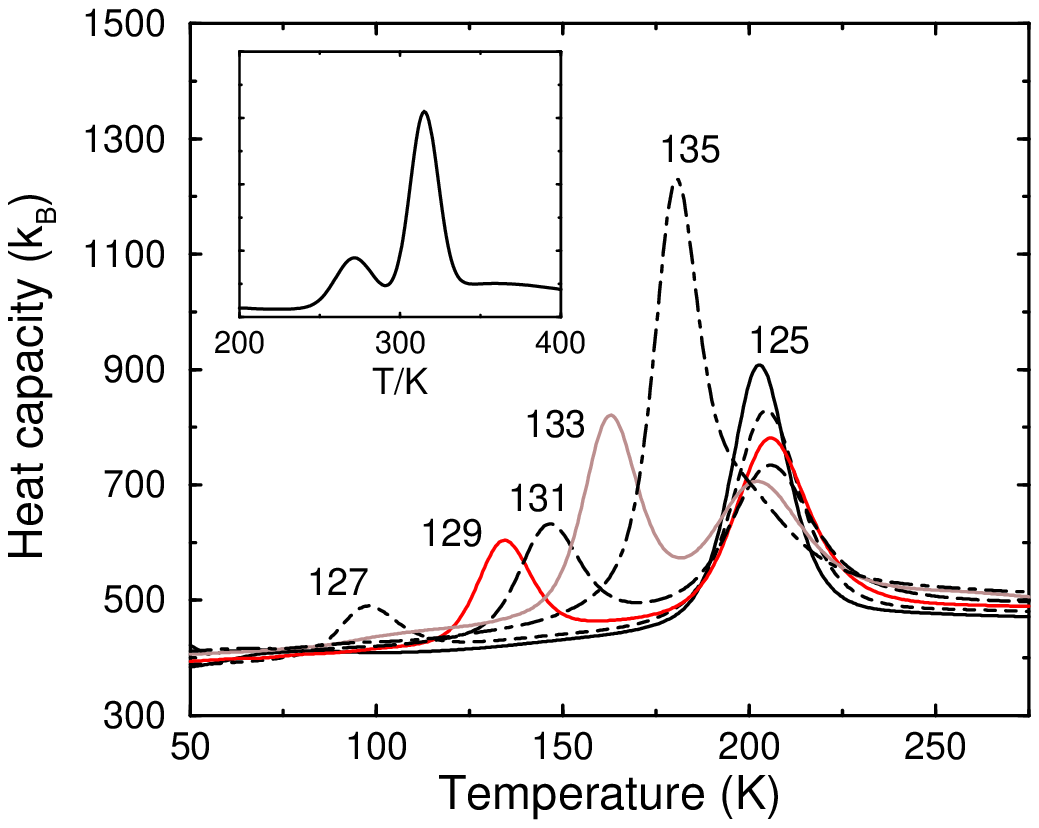}}
\caption{Heat capacities of sodium clusters calculated with exchange
Monte Carlo using the empirical potential, in the size range $125\leq
n\leq 135$. The inset shows the curve obtained for Na$_{133}$ using
the tight-binding quantum model.}
\label{fig:cv}
\end{figure}

The heat capacity curves in the range $125\leq n\leq 135$ are all
plotted in Fig.~\ref{fig:cv} versus temperature. Despite strong
changes from one size to another, a regular evolution can be seen
from 125 atoms and above this size. The heat capacity consists of
two peaks, the melting (or high-temperature) peak being centered
near 203~K. The smaller peak, denoted as premelting peak in the
following, goes from 100 to about 180~K in a quite continuous
fashion. The premelting peak is surprisingly strong, and can be
clearly distinguished from the melting peak. Strikingly, it is even
stronger than the melting peak itself at size 133.

\begin{figure}[htb]
\setlength{\epsfxsize}{9cm} \centerline{\epsffile{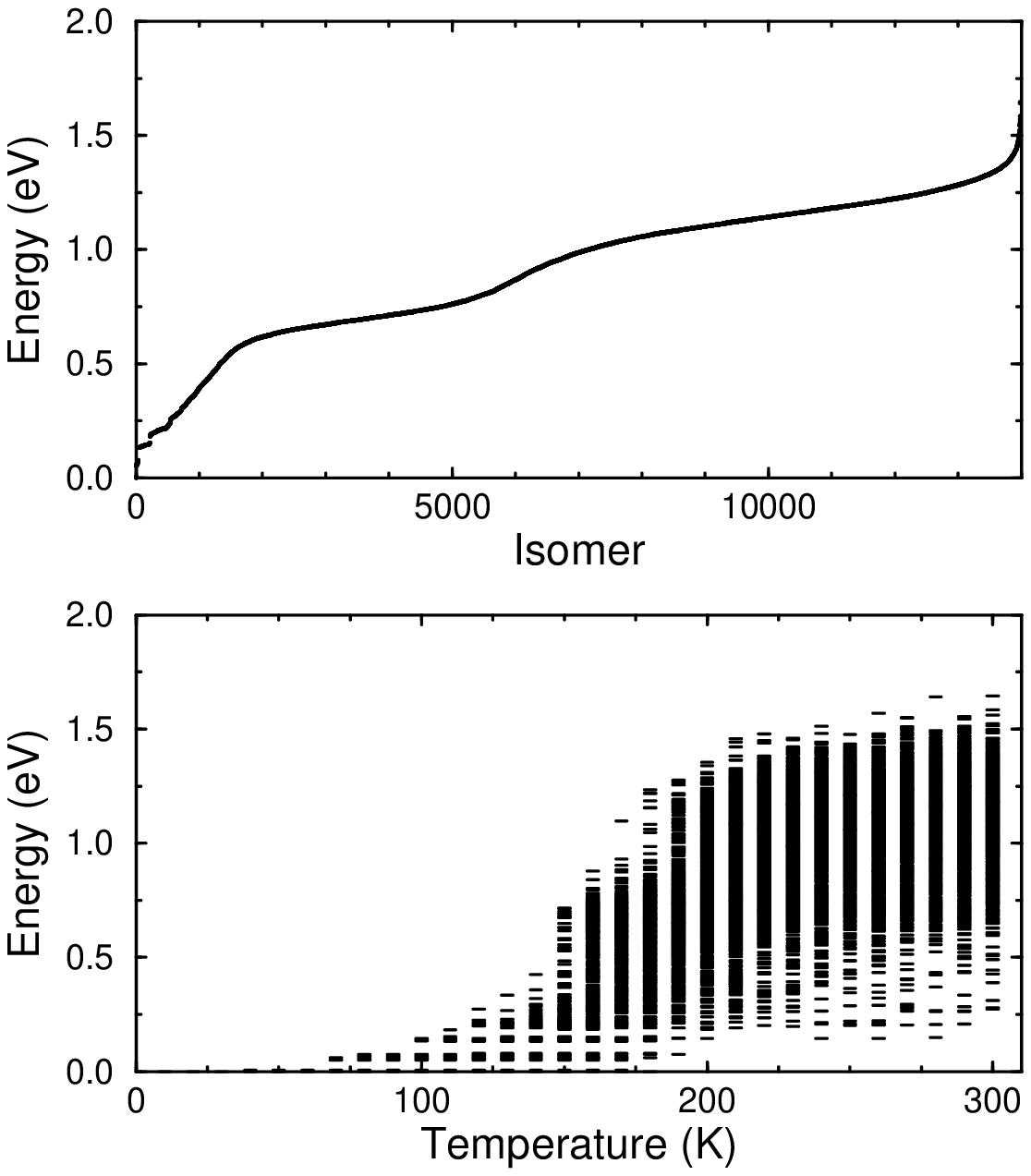}}
\caption{Quenching analysis of the Monte Carlo trajectories for
Na$_{133}$. Lower panel: spectra of isomers versus temperature.
Upper panel: energy versus isomer rank.}
\label{fig:133}
\end{figure}

To interpret these curves, we have chosen to focus on Na$_{133}$, by
carrying periodic quenches from instantaneous configurations extracted
from the Monte Carlo simulations for all trajectories. We thus
gathered nearly 14000 different isomers. Fig.~\ref{fig:133} shows the
energy of these isomers versus their rank, as well as the discrete
spectrum of isomers versus the temperature of the trajectory from which
they were quenched. Both graphs clearly indicate a correlation between
the repartition of isomers in energy space, their number and the heat
capacities. Small variations in $C_v$ occur below 150~K, but are
hardly visible when compared to the main peaks. They are related to
the first few hundreds of isomers, which involve the migration of a
limited number of some of the missing atoms on the external layer of
the icosahedron (the 135-atom cluster has $I_h$ symmetry with all
vertex atoms missing).

The presence of two major peaks in $C_v$ is consistent with the two main
increases in the number of isomers having less than a given
energy. Most new isomers appearing between 150 and 190~K have their
rank between 1000 and 5000 in the upper part of Fig.~\ref{fig:133}. 
Looking at their structure reveals that they are all still based on
the two-layer icosahedron, but that the third layer is hardly
recognisable. Hence this case of premelting is an extreme illustration
of surface melting following preliminar surface reconstruction.\cite{doyesurf}
Eventually, above 200~K the icosahedral structure is completely lost
and the true (volume) melting takes place.

If now we look more closely at the results obtained for Na$_{135}$, we
notice that the main peak is indeed located at the same temperature as
the premelting peak in Na$_{133}$, and that the true melting peak of
the latter cluster has been replaced by a right shoulder. Thus the
same phenomena seem to be present in Na$_{135}$, only with different
relative magnitude. This case could be referred to as 'post-melting'.

To some extent, this progressive evolution of the premelting and
melting peaks can be compared to what has been observed in small
Lennard-Jones (LJ) clusters by Frantz.\cite{frantz} In these systems, 
a premelting peak starts appearing close to the size 31 corresponding
to the competition between Mackay-type and anti-Mackay-type
icosahedral structures. The premelting peak remains as the cluster
size increases, and it is shifted to higher temperatures before
becoming higher than the melting peak itself near the size
38.\cite{frantz} Because of the structural similarities between LJ
clusters and the present sodium clusters, it is likely that the
present observations express the same qualitative mechanisms.
However, a significant quantitative difference can be seen in the
caloric curves of LJ clusters and sodium clusters, as the melting peak
is never really well resolved for van der Waals systems until it has
replaced the premelting peak. Here the two peaks are of very similar
widths, but their respective heights vary strongly, their total
contribution to the latent heat being nearly constant. 

Up to now, experimental data did not find any evidence for any
pronounced premelting peak in the heat capacities of charged sodium
clusters.\cite{hab1,hab2,hab3} This indicates that the present
empirical potential is {\em qualitatively}\/ unadequate to describe
these systems. In particular, beyond a simple scaling of the
parameters of the potential, we do not expect the use of explicit,
size-dependent parameters to improve the situation notably.
We repeated the above simulation for Na$_{133}$ using the more realistic
quantum distance-dependent tight-binding (TB) Hamiltonian described in
Ref.~\onlinecite{poteau}, but we had to reduce the statistics to
$10^6$ cycles following $2\times 10^5$ equilibration cycles (per
trajectory) for the computation to be tractable.
Even though melting points were shown to be
overestimated,\cite{cs2} we also noticed\cite{cs3} that premelting
effects were quite reduced using this model, in better consistency
with experiments. The heat capacity computed from exchange Monte Carlo
simulations is reported in the inset of Fig.~\ref{fig:cv}. For this
calculation we neglected the (weak) effects of nonzero electronic
temperature.\cite{cs3} The starting configuration was taken as the
same one as for the classical potential, but we did not find any more
stable structure during the course of the simulation.

At first sight, the caloric curve looks similar to the classical
result, with a clear premelting peak. However using the TB
Hamiltonian has two consequences. First, the premelting peak is much
lower than the melting peak, which is in agreement with our previous
general observation that the empirical potential overemphasizes
premelting features.\cite{cs2,cs3} More importantly, premelting also
occurs much closer in temperature to the melting peak itself. This
also suggest that premelting is not seen experimentally simply because
it is too broad.

\section*{Conclusion}

In the present work, we have obtained some evidence that premelting
effects in the caloric curves of sodium clusters could be present at
unexpectingly large sizes. We also found that some clusters could
exhibit 'post-melting', a process in which the premelting effect is
stronger than the actual melting peak. In the cases studied here,
these effects seem to be associated with surface reconstruction of the
third icosahedral layer, and thus seem to be of character similar to
what occurs in Lennard-Jones clusters having about 35 atoms.\cite{frantz}

One consequence of the above results is that 
explicit empirical potentials are not fully reliable for predicting
melting points in small sodium clusters, not only because they do not
allow one to reproduce the complex variations observed by the group
of Haberland,\cite{hab2} but mostly because they exhibit prominent
premelting peaks not seen in experiments. 

Calculations performed using the quantum tight-binding model also
predict a premelting phenomenon near 133 atoms, but the corresponding
anomaly of the heat capacity is much smaller than the melting peak,
as well as closer to it. In this respect, it resembles more the
measurements by Schmidt {\em et al.}\cite{hab2} 

Even though our calculations overestimate premelting effects,
they provide insight into the possible causes for the nonmonotonic
variations of the melting point. In particular, they suggest that
such variations may reflect premelting becoming actual melting. The
discrepancies with the present work would then be ascribed to a
possible merging of the premelting feature into one shoulder of the
melting peak, but not necessarily on the low temperature side.

\section*{Acknowledgments}

The authors wish to thank H. Haberland for helpful discussions.


\begin{thebibliography}{99}
\bibitem{difres} The q-jumping distributions obtained in the work of
Ref.~\protect\onlinecite{cs2} appeared to be incorrect due to loss of
numerical precision caused by the very small value of the Tsallis parameter
$q=3/2$, which should have been proportional to $n$.
\end{thebibliography}
\end{document}